# Simulation of breast compression using a new biomechanical model

Anna Mîra[{1,2}]*, Yohan Payan[2], Ann-Katherine Carton[1], Pablo Milioni de Carvalho[1], Zhijin Li[1], Viviane Devauges[1] and Serge Muller[1]

[1] GE Healthcare, Buc, F-78530;
[2]Univ. Grenoble Alpes, CNRS, Grenoble INP, VetAgro Sup, TIMC-IMAG, 38000 Grenoble, France

**ABSTRACT**

Mammography is currently the primary imaging modality for breast cancer screening and plays an important role in cancer diagnostics. A standard mammographic image acquisition always includes the compression of the breast prior x-ray exposure. The breast is compressed between two plates (the image receptor and the compression paddle) until a nearly uniform breast thickness is obtained. The breast flattening improves diagnostic image quality[1] and reduces the absorbed dose[2]. However, this technique can also be a source of discomfort and might deter some women from attending breast screening by mammography[3,4]. Therefore, the characterization of the pain perceived during breast compression is of potential interest to compare different compression approaches.

The aim of this work is to develop simulation tools enabling the characterization of existing breast compression techniques in terms of patient comfort, dose delivered to the patient and resulting image quality. A 3D biomechanical model of the breast was developed providing physics-based predictions of tissue motion and internal stress and strain intensity. The internal stress and strain intensity are assumed to be directly correlated with the patient discomfort. The resulting compressed breast model is integrated in an image simulation framework to assess both image quality and average glandular dose. We present the results of compression simulations on two breast geometries, under different compression paddles (flex or rigid).

## 1 INTRODUCTION

Today, mammography is the primary imaging modality for breast cancer screening and plays an important role in cancer diagnostics. Screening with mammography has been shown to reduce breast cancer mortality in women over 40-50 years in age. However, some women are still refusing to undergo regular mammography exams. Fleming et al[4] have shown, in a study of 2500 women, that 15% of those who skipped the second appointment, cited an unpleasant or painful first mammogram. The scope of this work is to compare the pain perceived by the women when their breasts are compressed using different designs of compression paddles, and therefore to identify the best compression paddle that may help improving the patient comfort and therefore the adherence to breast cancer screening.

The scope of this work is to assess the differences in terms of patient comfort, image quality and patient dose exposure when using a rigid or a flex paddle. Several studies showed that, the pain experienced by women during the mammographic exam depends on psychologic factor[5] (technician behavior, patient anxiety), sociologic factors[6] (ethnicity, education level) as well as physiologic[7] factors (compression level, breast size). Here, the psychologic and sociologic factors are neglected. The study focuses on physiological factors as the compression force or structural specifications of the compression paddle to characterize the patient comfort.

In this purpose, MR images of two subjects are used to create patient specific finite elements breast models. The mechanical behavior of soft tissues under compression is computed for both subjects and for both paddle designs. The perceived pain for a given paddle design is quantitatively characterized by contact pressure, internal stress and strain distributions. After compression, three sets of macrocalcifications are inserted into breast volumes. The latter are then subject to a Monte-Carlo based simulation (CatSim[8]) enabling to simulate the image acquisition of the compressed breast with a mammography system. Then, the diagnosis quality is assessed by measuring the signal-difference-to-noise-ratio (SDNR), signal-to-noise-ratio (SNR) and the average glandular dose (AGD).

## 2 MATERIALS AND METHODS

### 2.1 Subject specific data

To assess the entire breast volume as well as the surrounding soft tissues, the patient specific data is acquired using a specific MRI coil for breast imaging. The two volunteers taking part to this study agreed to participate in an experiment part of a pilot study approved by an ethics committee (MammoBio MAP-VS pilot study). The volunteers are 59 and 58 years old and have A-cup (volunteer 1) and F-cup (volunteer 2) breast size respectively. The images were acquired with a Siemens 3T MRI scanner using T2 weighted image sequences (Figure 1).

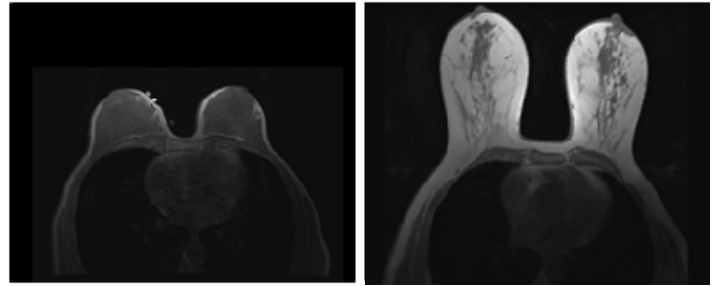

Figure 1 : Patient breast MR images in prone body position:
left - volunteer 1; right - volunteer 2.

The in-plane image resolution was 0.5x0.5 mm, and the slice thickness was 0.6 mm. The volunteers were also asked to provide the compression force and breast thickness as measured on their most recent mammograms. Such data are summarized in Table 1.

|  | Volunteer 1 | | Volunteer 2 | |
| --- | --- | --- | --- | --- |
|  | Right breast | Left breast | Right breast | Left breast |
| Force (N) | 21.9 | 40.9 | 94.8 | 56.6 |
| Breast thickness (mm) | 47 | 42 | 50 | 49 |

Table 1: Compression force and breast thickness for both volunteers for a cranio-caudal mammogram.

### 2.2 Biomechanical breast model

Breast deformation under compression is computed using the improved version of our previously developed breast biomechanical model[8]. The model shows a good fidelity to breast anatomy considering breast heterogeneity, anisotropy, personalized hyper-elastic properties of breast tissues and sliding boundary conditions between breast and pectoral muscle.

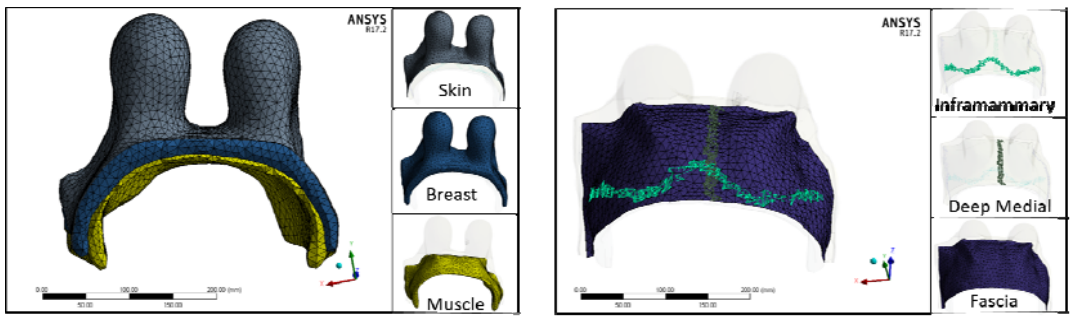

Figure 2: Biomechanical breast model: left panel – breast components; right panel breast suspensory matrix.

The 3D geometrical breast shape is generated from segmented MR images of the volunteer using ITK-snap[6] software. The 3D breast geometry is discretized in a tetrahedral Finite Element (FE) mesh and is subject to a hyper-elastic quasi-static deformation using the ANSYS software framework[7]. Five types of tissues are considered: muscle, adipose and glandular tissues (here referred as breast tissues), skin, fascia and ligaments (Figure 2).

The muscle, breast tissues, skin and fascia are modeled as homogeneous, hyper-elastic quasi-incompressible materials (Poisson ratio $\upsilon = 0.499$). Mechanical response of soft tissues to external stimuli was modeled using the Gent form of strain-energy potential[9]. Considering small strain behavior, the Gent expression reduces to the Neo-Hookean one[10]. Therefore the Young's moduli (E) of the tissues were chosen in conformity with the Neo-Hookean constitutive models described in our previous work[8] and are summarized in Table 2. The limiting value $J_m$ of the first invariant of the left Cauchy-Green deformation tensor was fitted to comply with the compression data from Table 2.

|  | $E_{muscle}$ (kPa) | $E_{breast}$ (kPa) | $E_{skin}$ (kPa) | $E_{fascia}$ (kPa) | $J_m$ |
|---|---|---|---|---|---|
| Subject 1 | 10 | 0.4 | 3 | 120 | 1 |
| Subject 2 | 10 | 0.5 | 10 | 160 | 2 |

Table 2: Constitutive parameters for the used materials: E - Young's modulus; Jm - the limiting values of the first invariant of the left Cauchy-Green deformation tensor.

## 2.3 Compression paddles

During mammography, a qualified radiologic technologist positions the breast of the patient between a stationary image receptor and a movable paddle (Figure 3). The technologist gradually compresses the breast in order to even out the breast thickness and to spread out the soft tissues. Nowadays, two types of compression paddles are widely available: rigid compression paddles (RCP) and flex compression paddles (FCP).

The RCP is fixed to its frame and is constrained to move in the up-down direction. This paddle has some flexibility because of material mechanical properties and can slightly bend when compressing the breast, while remaining globally flat and parallel to the image receptor. On the other hand, the FCP is attached to its frame by rotational joints and therefore, presents an additional rotational degree of freedom enabling the paddle to tilt with respect to the image receptor plane (Figure 3.c).

With a rigid paddle, the breast under compression presents a nearly uniform thickness all over the contact surface. Contrariwise, with a flex paddle, the compressed breast thickness decreases quasi linearly from the chest wall to the nipple. Flex paddles are used to better conform the breast contours and thereby to improve compression. However, Broeders and collegues[11] have shown that such compression paddle may decrease the diagnostic quality of mammograms as the breast tissues may be pushed out to the chest wall resulting in less retro-glandular tissue visible on the image.

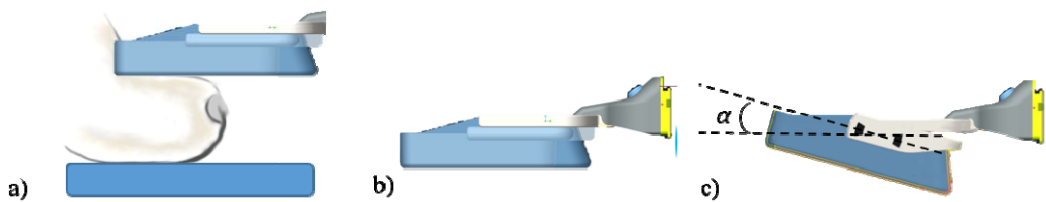

Figure 3: Breast compression between the paddle (up) and the receiver (down): a) Rigid paddle; b) Flex paddle with flexion angle α.

## 2.4 Breast compression

The prone configuration of the breast is used to simulate the effect of the breast pulling gesture performed by the technologist during breast positioning. Neglecting the gravity effects, the breast geometry is then tilted to the body upright position. To simulate the cranio-caudal incidence, the image receptor is positioned at the inframammary ligament level while the paddle compresses the breast by a downward movement. The minimal thickness and maximal compression force were consistent with the subject real acquisition data (Table 1).

A finite element model was defined for each type of compression paddles based on the geometry specifications from a Senographe Pristina$^{TM}$ mammography unit (GE-Healthcare, Buc, France). On both cases, the paddle flexibility due to the material properties was neglected. The RCP was therefore defined as a fully rigid body with only one translational degree of freedom in the downwards direction. For the FCP, a rotation around the longitudinal axis is added. An additional degree of freedom was modeled using a rotational-only joint type of element. The joint stiffness was computed by fitting the force-deflection curve of a flex paddle from a Senographe Pristina$^{TM}$ unit.

The compression was then computed for both right breast volumes. To compare the two types of compression, a nominal breast thickness was defined. The nominal thickness for the RCP was defined as the distance between the paddle and the image receptor. The nominal thickness for the FCP was defined as the distance between the compression paddle and the image receptor at mid-length of the compressed breast. Then, we compared the contact pressure and the internal strain/stress distributions. The nominal breast thickness was computed and compared at different compression levels.

## 2.5 Image quality and average glandular dose

To assess the impact of breast compression on image quality, we inserted a set of microcalcifications into each compressed breast volume. The smallest breast volume contains 21 microcalcifications arranged in a matrix of 7 rows and 3 columns (Figure 4a). The largest breast volume contains 56 microcalcifications arranged in a matrix of 7 rows and 8 columns. The matrix of calcifications is parallel with the entrance surface of the image receptor and positioned at the breast mid thickness (Figure 4b). The distance between two consecutive columns or rows is 10mm. We assumed a uniform breast-equivalent material composed of glandular/adipose tissue with a 20/80 ratio. A mammogram was simulated using typical clinical acquisition parameters obtained with the standard automatic optimization of parameters (AOP) mode. Two simulations were performed with microcalcifications of 0.2 mm and 0.3mm in diameter. The signal-difference-tonoise ratio (SDNR) per pixel of these microcalcifications was measured. Additionally, the signal-to-noise ratio (SNR) was computed on the same pixels excluding the microcalcifications.

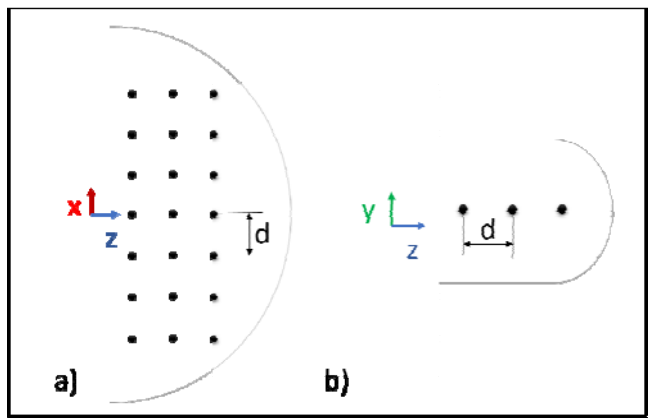

Figure 4: Microcalcification distribution over the smallest breast volume (d = 10mm): a) axial view, b) sagittal view.

The average glandular dose (AGD) was derived using the approach proposed by Dance et al[12] regardless the paddle type. In practice, it is very difficult to accurately measure the exact breast thickness. Thus, the nominal breast thickness was used to compute conversions factors which relate measurements of incident air kerma to the delivered mean glandular dose.

# 3 RESULTS AND DISCUSIONS

The force versus breast thickness curves are plotted in Figure 5 for both volunteers with the rigid and flex paddles. We observed a nominal compression thickness roughly equals for both paddles. The resulting internal stress and strain distributions, as well as contact pressure maps were derived at compressive forces of 22 N for the first volunteer (Figure 6) and 95 N for the second one (Figure 7).

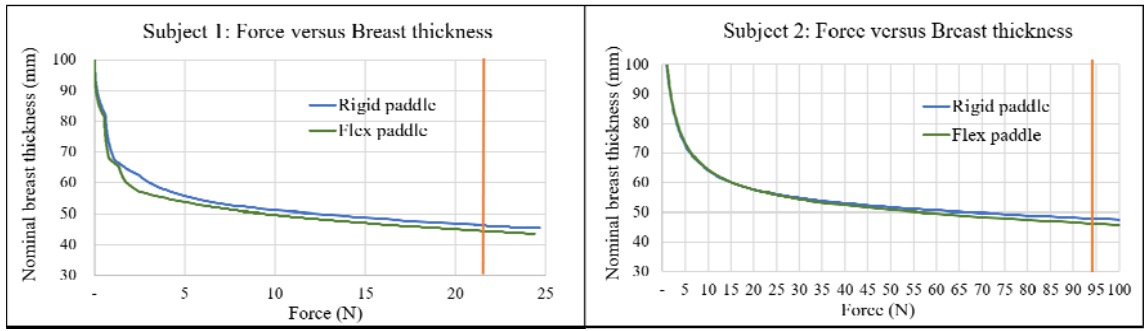

Figure 5: Resulting breast thickness for a given compression force

As concerns the small breast volume (Figure 6), there is no significant difference between FCP and RCP in pressure distribution over the skin surface or in internal stress/strain intensity distributions. For both compression paddles, high pressure at the skin surface is concentrated in the juxtathoracic region with a maximum pressure of 77.7 kPa. Several clinical studies[11,13] sustained this result of no significant difference in experienced pain when using FCP or RCP. In addition, the FE simulations confirm that in small breasts the paddle tilt is too small to impact the tissues compression in the middle part of the breast.

FCP applied on large breast volumes (Figure 7) results in significantly lower intensities of pressure at the skin surface in contact with the compression paddle, with a maximal pressure of 37 kPa, compared to 56 kPa when using RCP. No significant difference in the measured maximal intensities of strain and stress was observed, however strain and stress distribution patterns are different. When the breast is compressed with a rigid paddle, maximal strain and stress are concentrated in the retromammary space and decrease considerably toward the nipple. When a flex paddle is used, stress and strain are more uniformly distributed over the breast volume with the highest values in the middle third of the breast.

The areal pressure distribution patterns has already been demonstrated in the work by Dustler and collegues[13]. The authors have studied the pressure distribution patterns of 103 women undergoing breast compression with a rigid paddle at different compression levels. Four groups have been differentiated: a) skin pressure widespread over the breast (29%); b) skin pressure concentrated on the central part of the breast (8%); c) skin pressure concentrated on the juxtathoracic region (16%); d) skin pressure concentrated along a narrow zone at the juxtathoracic region (26%). The pressure distribution patterns observed for our first and second volunteers correspond to the group d and a respectively.

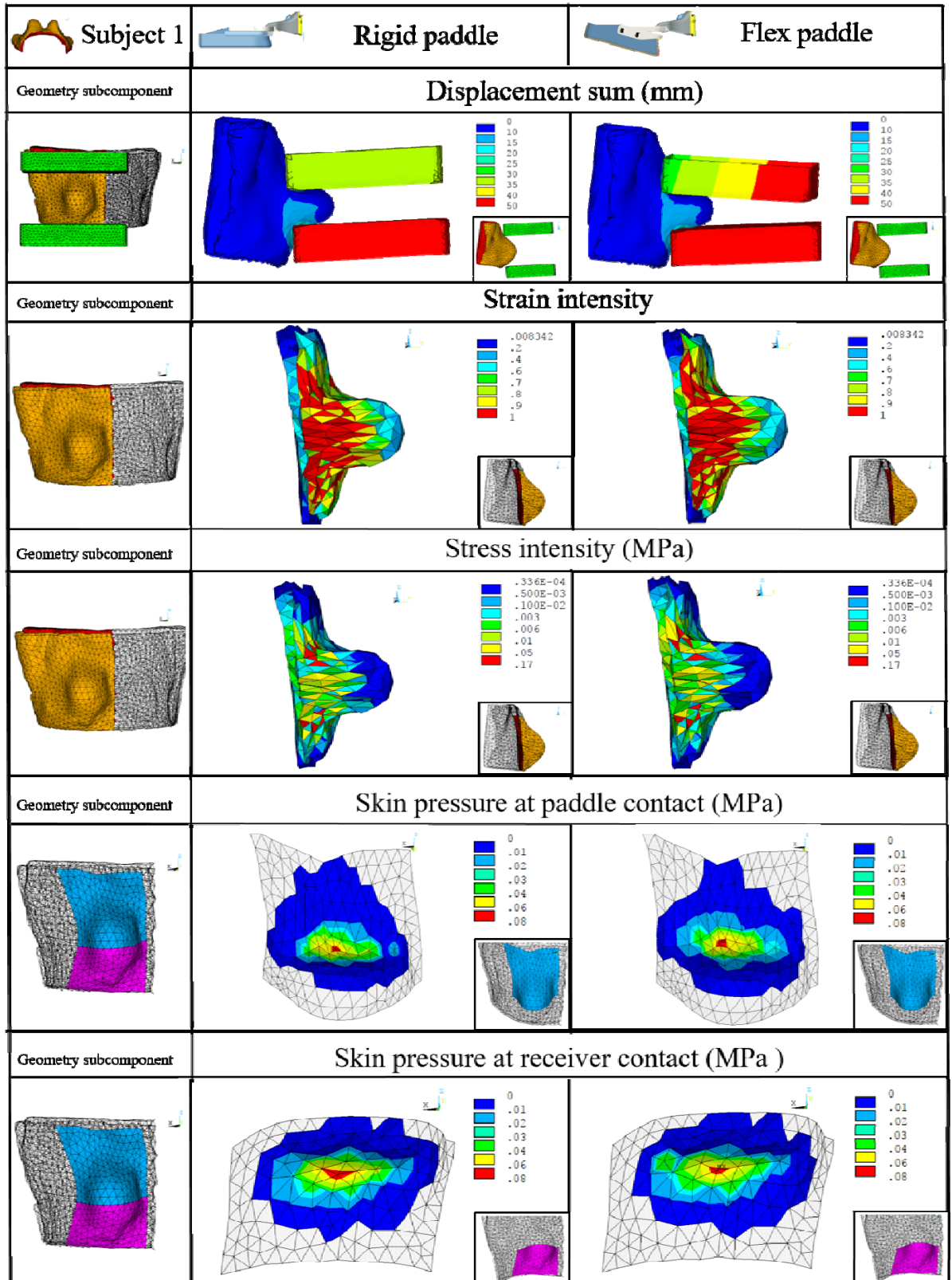

Figure 6: Stress, strain and contact pressure distribution for the first subject.

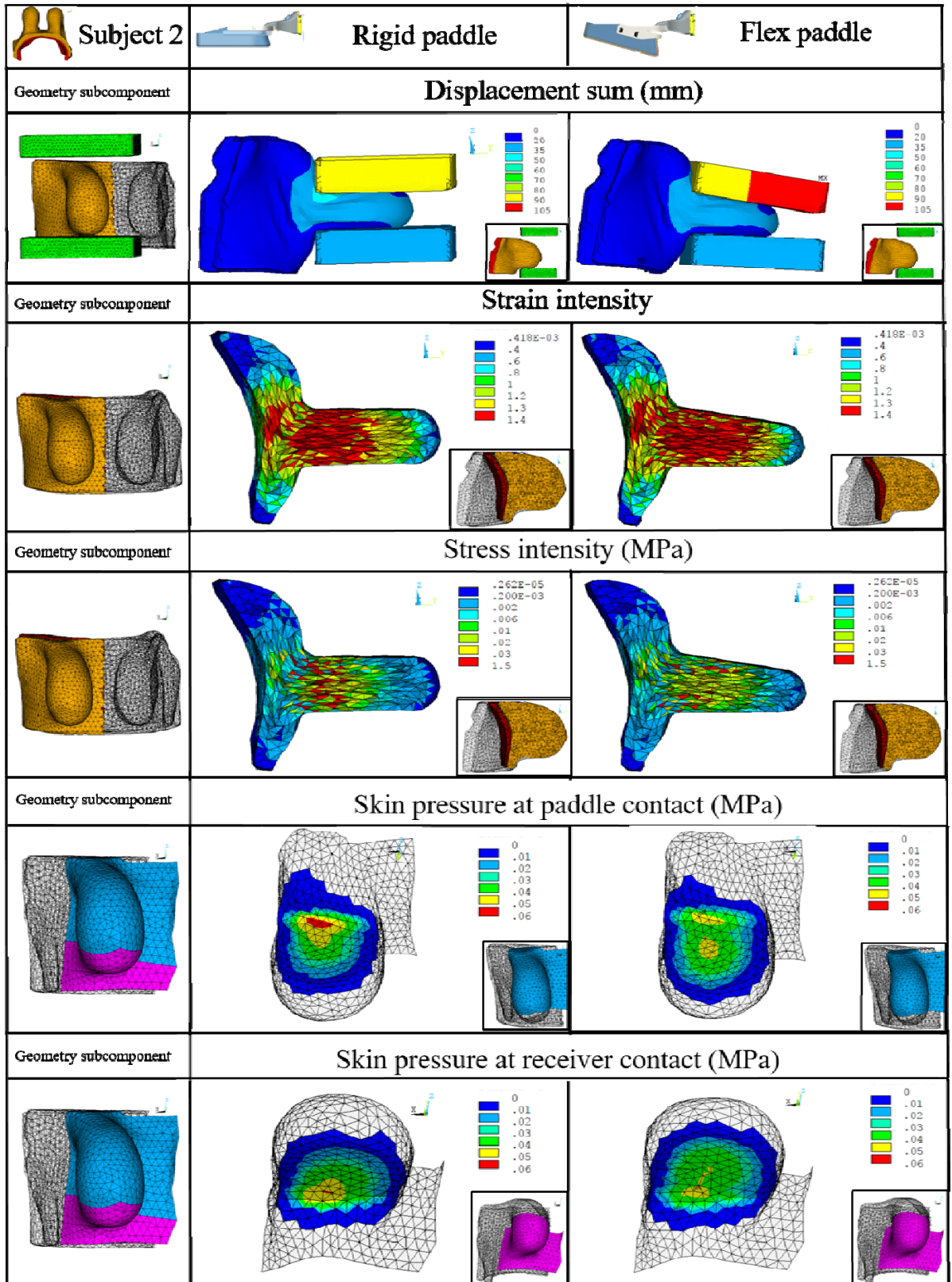

Figure 7: Stress, strain and contact pressure distribution for the second subject.

The nominal breast thickness may vary by about 2mm between rigid and flex paddle for both volunteers (Table 3). Accordingly, no significant difference was found between the estimated AGD, while a dose reduction of 2% for the smaller breast and 4% for larger breast was observed.

|  | Rigid Paddle | | Flex Paddle | | |  | Rigid Paddle | | Flex Paddle | |
|---|---|---|---|---|---|---|---|---|---|---|
|  | Mean SNR | StdDev SNR | Mean SNR | StdDev SNR | p-Values |  | BNT (mm) | AGD (mGy) | BNT (mm) | AGD (mGy) |
| Volunteer 1 | 82,90 | 43,09 | 83,70 | 37,72 | 0,706 |  | 46 | 1,15 | 44 | 1,12 |
| Volunteer 2 | 126,89 | 8,75 | 137,21 | 10,73 | 0,000 |  | 48 | 1,20 | 46 | 1,15 |

|  | 200 um | | | | | 300 um | | | | |
|---|---|---|---|---|---|---|---|---|---|---|
|  | Rigid Paddle | | Flex Paddle | | | Rigid Paddle | | Flex Paddle | | |
|  | Mean SDNR | StdDev SDNR | Mean SDNR | StdDev SDNR | p-Values | Mean SDNR | StdDev SDNR | Mean SDNR | StdDev SDNR | p-Values |
| Volunteer 1 | 0,74 | 0,68 | 0,79 | 0,54 | 0,689 | 2,01 | 1,28 | 1,85 | 1,02 | 0,224 |
| Volunteer 2 | 1,14 | 0,57 | 1,13 | 0,53 | 0,885 | 2,96 | 0,76 | 3,15 | 0,92 | 0,093 |

Table 3: Breast nominal thickness (BNT), average glandular dose (AGD), signal-to-noise-ratio (SNR) and signal-difference-to-noise (SDNR) for both volunteers and both compression paddle types.

The SNR and SDNR have been estimated and compared between flex and rigid paddles. When using a flex paddle instead of a rigid paddle on the largest breast (volunteer 2), we observe a statistically significantly higher SNR. The same trend is observed on SDNR for both 200 and 300 μm microcalcifications, while not statistically significant. We did not observe any statistically significant difference in SNR or SDNR for microcalcification of any size when considering the compression of the smallest breast by a rigid or a flex paddle. Therefore, despite a breast thickness varying linearly from chest wall to nipple when the flex compression paddle is used, the image quality is preserved or improves compared to the image quality obtained with the rigid compression paddle.

## 4 CONCLUSION

Breast compression with flex and rigid paddle have been simulated using the finite elements theory applied to segmented MRI images acquired on 2 volunteers under different geometries. Appling the Gent form of strain-energy potential, instead of the Neo-Hookean form, allowed to obtain compression force magnitudes comparable with the real subject data.

After simulating the breast compression, the SDNR of microcalcifications and the AGD, delivered during the acquisition of the corresponding simulated mammography, have been computed. The simulations have been repeated for two different breast volumes (cup sizes A and F) with a rigid and a flex paddle. The four configurations have been analyzed to compare patient perceived pain (measured as strain and stress) and image quality (measured as SNR, SDNR and AGD). The results of our simulations indicate that, for the smallest breast, there is no significant difference for the patient perceived pain when using the rigid or the flex paddle. The shape of the breast under compression does not present significant changes between the two paddle designs. We did not observe any statistically significant difference in SNR or SDNR for microcalcification of any size when considering the compression of the smallest breast by a rigid or a flex paddle. Therefore, our results suggest that using a flex paddle should not significantly impact image quality and delivered dose in small breasts, and should not reduce significantly the perceived pain.

For the largest breast, our simulations indicate that using a flex paddle may reduce the maximal pressure intensity on the skin surface by about 30% compared to the rigid paddle. The tissues deformation is more uniformly distributed inside the breast volume, and the highest deformation is occurring in the middle breast region corresponding to the supposed location of dense tissues. Moreover, our simulations have shown that flex paddle have no significant impact on the average glandular dose and improves image quality compared to the rigid paddle.

In conclusion, our simulations confirm that using the flex paddle used for breast compression may improve the patient comfort without affecting the image quality and the delivered average glandular dose. Moreover, despite a breast thickness varying linearly from chest wall to nipple, when a flex compression paddle is used on large breasts, the image quality seems to be preserved or improved compared to the image quality obtained with a rigid compression paddle.

# REFERENCES

1.	Saunders Jr, R. S., Samei, E., Lo, J. Y., & Baker, J. A., "Can compression be reduced for breast tomosynthesis? Monte Carlo study on mass and microcalcification conspicuity in tomosynthesis." *Radiology*, *251*(3), 673-682 (2009).
2.	Chen, B., Wang, Y., Sun, X., Guo, W., Zhao, M., Cui, G., ... & Yu, J., "Analysis of patient dose in full field digital mammography." *European journal of radiology*, *81*(5), 868-872 (2012).
3.	Aro, A. R., De Koning, H. J., Absetz, P., & Schreck, M., "Psychosocial predictors of first attendance for organised mammography screening." *Journal of Medical Screening*, *6*(2), 82-88 (1999).
4.	Fleming, P., O'Neill, S., Owens, M., Mooney, T., & Fitzpatrick, P., " Intermittent attendance at breast cancer screening." *Journal of public health research*, *2*(2) (2013).
5.	Aro, A. R., Absetz-Ylöstalo, P., Eerola, T., Pamilo, M., & Lönnqvist, J. "Pain and discomfort during mammography." *European Journal of Cancer*, *32*(10), 1674-1679 (1996).
6.	Dullum, J. R., Lewis, E. C., & Mayer, J. A., "Rates and correlates of discomfort associated with mammography." *Radiology*, *214*(2), 547-552 (2000).
7.	Poulos, A., McLean, D., Rickard, M., & Heard, R., "Breast compression in mammography: how much is enough?" *Journal of Medical Imaging and Radiation Oncology*, *47*(2), 121-126 (2003).
8.	Milioni de Carvalho, P. "Low-dose 3D quantitative vascular X-ray imaging of the breast". PhD thesis. University Paris 11, (2014).
9.	Mîra, A., Carton, A. K., Muller, S., & Payan, Y. "Breast Biomechanical Modeling for Compression Optimization in Digital Breast Tomosynthesis." In *Computer Methods in Biomechanics and Biomedical Engineering,* 29-35 (2018).
10.	Yushkevich, P. A., Piven, J., Hazlett, H. C., Smith, R. G., Ho, S., Gee, J. C., & Gerig, G., "User-guided 3D active contour segmentation of anatomical structures: significantly improved efficiency and reliability." *Neuroimage*, *31*(3), 1116-1128 (2006).
11.	ANSYS®, R. Academic Research Mechanical, Help System, Modeling and Meshing Guide, Help System. (17.2).
12.	Gent, A. "A new constitutive relation for rubber." *Rubber Chem. Technol.* **69 (1),** 59--61 (1996).
13.	Chagnon, G., Marckmann, G., & Verron, E. "A comparison of the Hart-Smith model with Arruda-Boyce and Gent formulations for rubber elasticity." *Rubber chemistry and technology*, *77*(4), 724-735 (2004).
14.	Broeders, M. J., ten Voorde, M., Veldkamp, W. J., van Engen, R. E., van Landsveld–Verhoeven, C., NL't Jong–Gunneman, M., ... & den Heeten, G. J., "Comparison of a flexible versus a rigid breast compression paddle: pain experience, projected breast area, radiation dose and technical image quality." *European radiology*, *25*(3), 821-829 (2015).
15.	Dance, D. R., & Young, K. C., "Estimation of mean glandular dose for contrast enhanced digital mammography: factors for use with the UK, European and IAEA breast dosimetry protocols." *Physics in medicine and biology*, *59*(9), 2127 (2014).
16.	Dustler, M., Andersson, I., Brorson, H., Fröjd, P., Mattsson, S., Tingberg, A., ... & Förnvik, D., "Breast compression in mammography: pressure distribution patterns." *Acta Radiologica*, *53*(9), 973-980 (2012).